\documentclass[twocolumn,twoside,slac_two]{revtex4}
\usepackage{graphicx}
\usepackage{fancyhdr}
\begin{document}

\title{Discovery Potential of the Standard Model Higgs Boson Through $H \rightarrow WW$ Decay Mode with the ATLAS Detector at LHC}

%

\author{H.J. Yang (on behalf of the ATLAS Collaboration)}
\affiliation{Department of Physics, University of Michigan, Ann Arbor, MI 48109-1120, USA}

\begin{abstract}
We report results of a study of the Standard Model Higgs boson discovery potential through
the W-pair leptonic decay modes with the ATLAS detector at LHC at 14 TeV center-of-mass energy. 
We used MC samples with full detector simulation and reconstruction of the ATLAS
experiment to estimate the ATLAS detection sensitivity for the reaction of
$pp \rightarrow H \rightarrow WW \rightarrow e\nu \mu\nu$ with no hard jet
or two hard jets in the final states. The prospects for the Higgs boson searches
at ATLAS are presented, including trigger efficiencies and data-driven methods 
to estimate the backgrounds using control samples in data.
With 10 fb$^{-1}$ of integrated luminosity, one would expect to discover a Standard
Model Higgs boson with ATLAS detector in Higgs boson mass range $135 < m_H < 190$ GeV.
If the Higgs boson does not exist, we will extend and confirm the exclusion produced by 
the Tevatron Higgs boson search result, which has ruled out the Higgs boson mass range of 
$160 < m_H < 170$ GeV at 95\% confidence level. 
If Higgs boson would be discovered, ATLAS could measure its mass
with a precision of about 2 and 7 GeV for Higgs boson mass around 160 GeV and 130 GeV, 
respectively.
\end{abstract}

\maketitle

\thispagestyle{fancy}


\section{Introduction}

The Standard Model (SM) of particle physics is a successful theory
confirmed by numerous measurements since the 1970's.
After the direct observations of the massive top quark and the
$\nu_\tau$ at Fermilab in 1995 and 2000, respectively, only one
fundamental particle predicted by the SM, the Higgs boson,
remains to be discovered. The role of the Higgs boson is critical in the SM
because fermions and W/Z bosons, via a process of spontaneous symmetry breaking,
acquire their mass through  interactions with the Higgs boson fields.
In the SM the Higgs boson couplings to all fermions and electroweak gauge bosons
depend on their masses. Generically, the SM Higgs boson couples most
strongly to heavy particles. However, the SM doesn't predict the Higgs boson mass,
which resulted in a great experimental challenging to search for the Higgs
boson in a very wide mass range and through many different decay modes.

\begin{figure}[h]
\begin{center}
\includegraphics [width=0.45\textwidth] {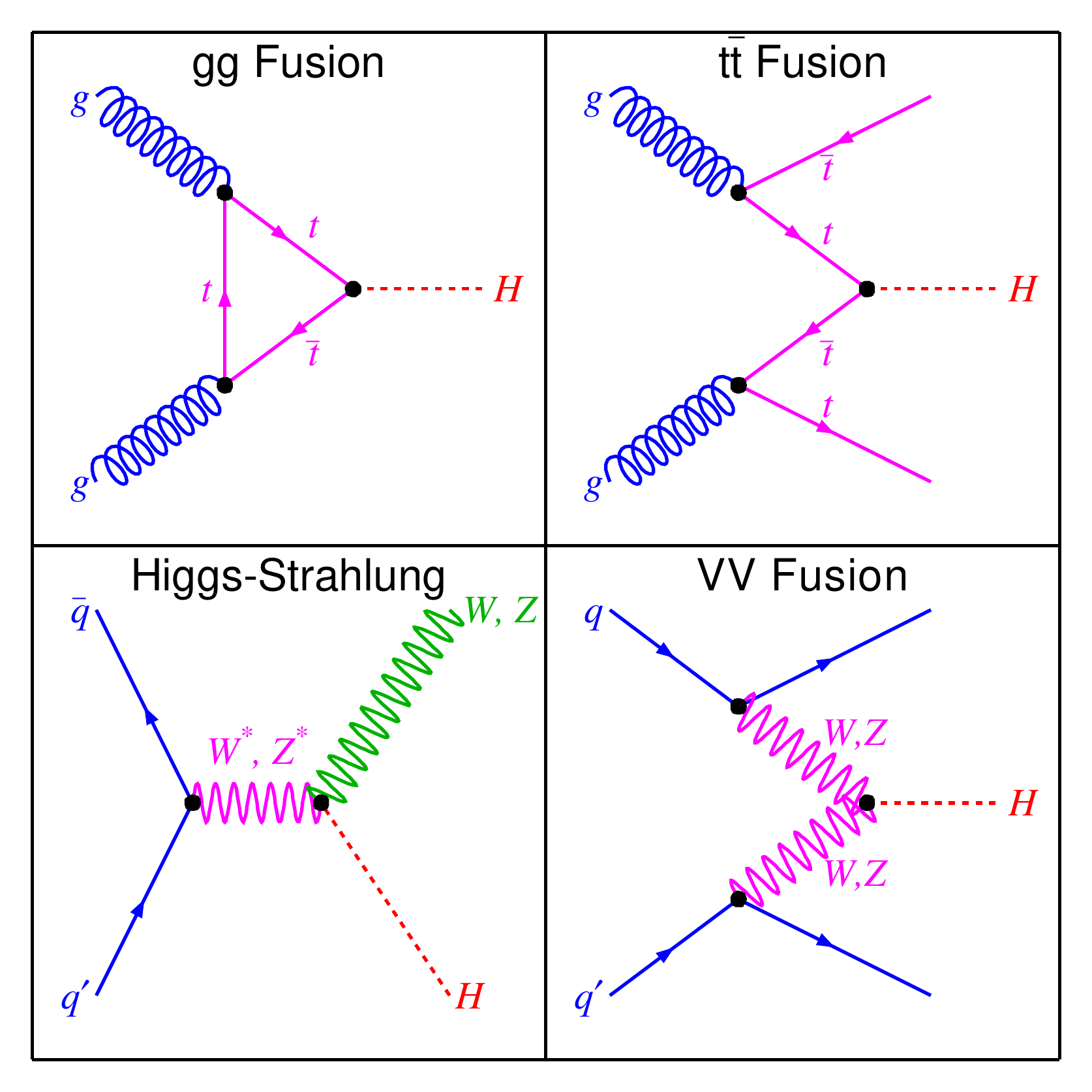}
\caption{SM Higgs boson production LO Feynman diagrams.}
\label{fig:higgs_feyn}
\end{center}
\end{figure}

\begin{figure}[h]
\begin{center}
\includegraphics [width=0.48\textwidth] {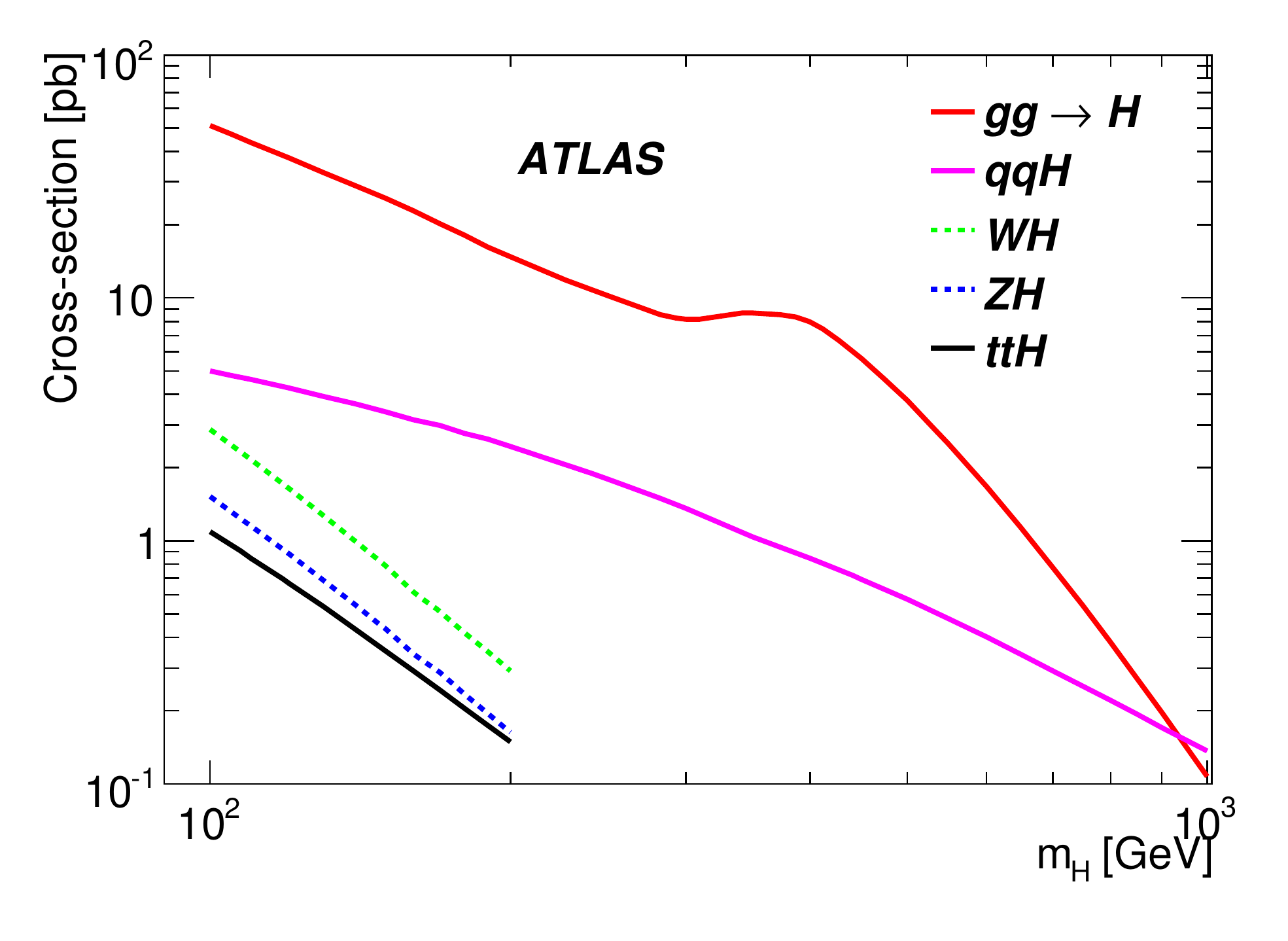}
\caption{SM Higgs boson production NLO cross-sections at the LHC at 14 TeV}
\label{fig:higgs_xsec}
\end{center}
\end{figure}

\begin{figure}[h]
\begin{center}
\includegraphics [width=0.45\textwidth] {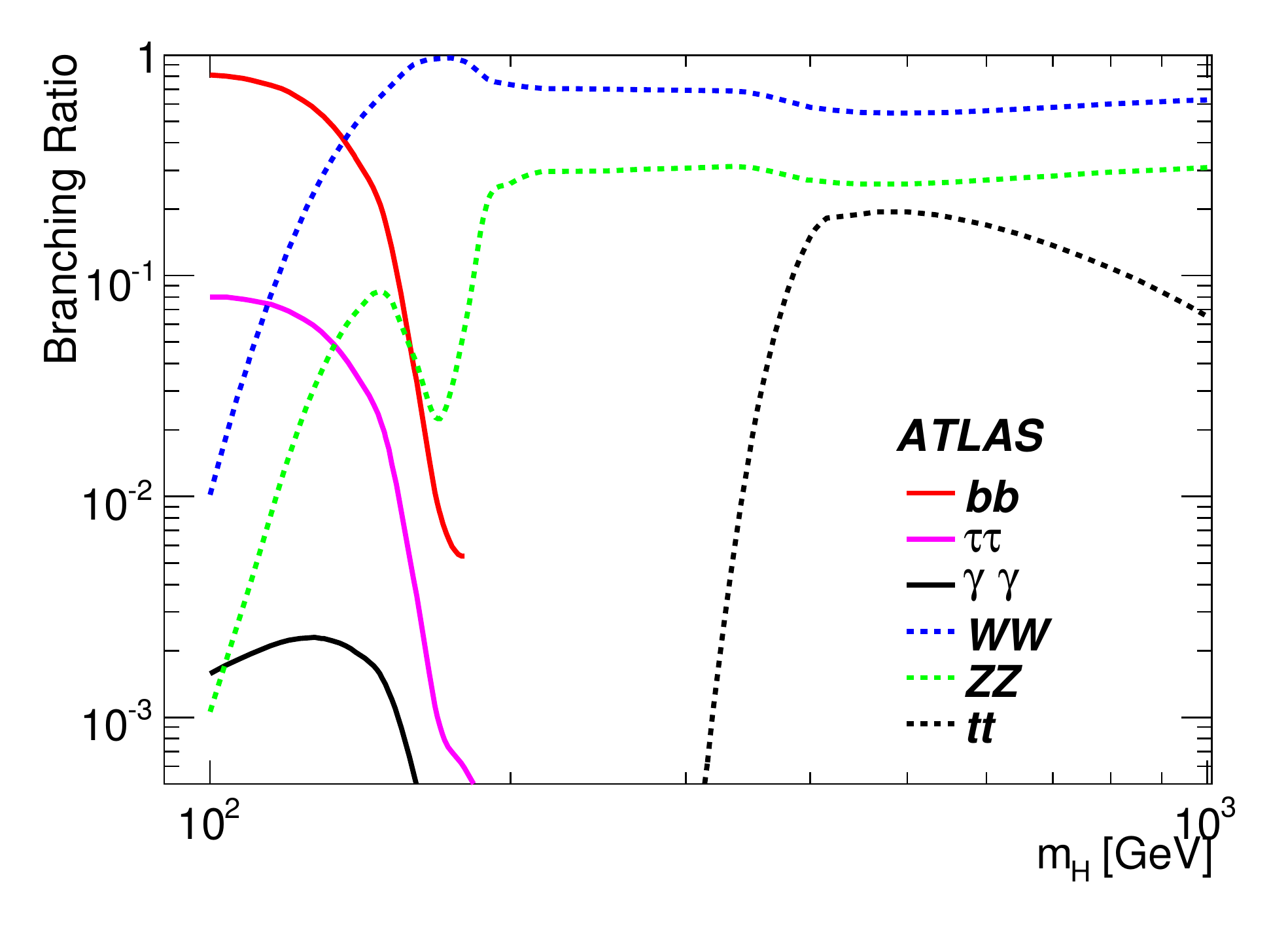}\hfill
\caption{SM Higgs boson decay branching ratios}
\label{fig:higgs_decay}
\end{center}
\end{figure}

The Leading Order (LO) SM Higgs boson production mechanism at the LHC (Large Hadron Collider) is shown
by the Feynman diagrams in Figure \ref{fig:higgs_feyn}, 
including the processes of the Gluon-Gluon Fusion (GGF), the $t\bar t$ fusion, the Higgs-strahlung and the Vector-Boson
Fusion (VBF). The dominant process is the GGF Higgs boson as shown in the next-to-leading-order (NLO) cross-section
plot in the Figure~\ref{fig:higgs_xsec}. Higgs boson production via
VBF is also an important production mechanism for Higgs boson searches at the LHC
with the specific signature of a Higgs boson with two hard jets in the final states. 
The Higgs boson production cross-section is very small compared to the enormous QCD jet productions at the LHC.
The ability to detect and identify the Higgs boson decay final states (see Figure \ref{fig:higgs_decay})~\cite{atlas}
is crucial to the success of the Higgs boson discovery.

The direct search at LEP sets a lower Higgs boson mass limit of 114.4 GeV at 95\% confidence level~\cite{lep}.
Recent searches at Tevatron exclude the Higgs boson mass between 160 GeV and 170 GeV at 95\% confidence level~\cite{tevatron}.
The electroweak data constraints deduced from consistency conditions
of the SM can be used to derive an upper limit of SM Higgs boson mass of ~190 GeV at
95\% C.L.\cite{atlas,cms}. Therefore the most interesting mass range indicated by
direct Higgs boson search and indirect constraints is between 114 GeV and 190 GeV.

\section{Monte Carlo Samples}

As shown in the Figure~\ref{fig:higgs_xsec}, there are two dominant Higgs boson production modes: GGF and VBF, 
which are in the kinematic region of interest for the $H \rightarrow WW$
decay mode. When both W bosons decay to leptons, the Higgs boson signature includes two energetic leptons with
large missing transverse energy in the final state. These two leptons tend to go same direction due to
spin correlation. For the VBF Higgs boson, the final state also includes two hard jets which tend to be
well-separated in pseudo-rapidity. The Higgs boson signal and major background sources (SM $WW$, $t\bar{t}$,
Z+jets and W+jets) are listed in Table~\ref{montecarlo}. The MC generators we used to 
produce MC events and corresponding cross sections are also listed in the Table~\ref{montecarlo}.

\begin{table}[h]
\begin{center}
\caption{Monte Carlo generators and cross sections for the Higgs boson signal and background processes.
The W+jets cross sections listed is the cross section per lepton flavor.}
\begin{tabular}{|l|c|c|}
\hline \textbf{Physics Process} & \textbf{Generator} & \textbf{$\sigma$(pb)} \\ \hline 

$gg \rightarrow H \rightarrow WW$($M_H=170$ GeV) & MC@NLO          &  19.418 \\ \hline
VBF $H \rightarrow WW$($M_H=170$ GeV)            & Pythia          &   2.853 \\ \hline
$qq/qg \rightarrow WW$                           & MC@NLO          & 111.6   \\ \hline
$gg \rightarrow WW$                              & GG2WW           &   5.26  \\ \hline
$pp \rightarrow t\bar{t}$                        & MC@NLO          & 833.0   \\ \hline
$Z \rightarrow \tau\tau$ + jets                  & ALPGEN          & 2015    \\ \hline
$W \rightarrow \ell\nu$ + jets                   & ALPGEN          & 20510   \\ \hline
\end{tabular}
\label{montecarlo}
\end{center}
\end{table}

\section{Leptonic W Pair Production with No Hard Jets}

The event selection for $H \rightarrow WW \rightarrow \ell\nu\ell\nu$ channel consists of 
a set of simple cuts.
\begin{itemize}
\item Require that the selected event has exactly two isolated, opposite-charge leptons (electron or muon)
with $P_T > 15$ GeV.
\item To suppress single-top background, backgrounds from dileptonic decays of $b\bar{b}$ and
$c\bar{c}$ resonances, and lepton pairs from $b \rightarrow c$ cascade decays, require that the
invariant mass $m_{\ell\ell}$ of the leptons is between 12 GeV and 300 GeV.
\item Require that the event has missing transverse energy $E_T^{miss} > 30$ GeV.
\item To suppress $Z \rightarrow \tau\tau$ background, reject events which have invariant mass 
of a hypothetical $\tau$ pair in the range $|M_{\tau\tau} - M_Z| < 25$ GeV.
\item To suppress backgrounds from top quark decays, reject events that contain any hard jets with
$p_T > 20$ GeV and $|\eta_j| < 4.8$.
\item To further suppress the top-related background, reject events with any jets with 
$p_T > 15$ GeV and a b-tagging weight greater than 4.
\end{itemize}

Table~\ref{cut_flow_h0j} shows the cross sections in the $e\mu$ channel for signal and background
after these cuts.
\begin{table*}[t]
\begin{center}
\caption{Cut flows(in fb) for $M_H=170$ GeV in the H + 0j, 
$H \rightarrow WW \rightarrow e\nu\mu\nu$ channel.
The WW background contains the two processes $q\bar{q} \rightarrow WW$ and $gg \rightarrow WW$.}
\begin{tabular}{|l||c||c|c|c|c|c|}
\hline\textbf{Selection} & Selection Cuts & $~~gg \rightarrow H$~~ & ~~$t\bar{t}$~~  & ~~ WW ~~ & ~~ $Z\rightarrow \tau\tau$ ~~ & ~~W+jets ~~\\ \hline
                  & Lepton Selection + $M_{\ell\ell}$ & 166.4 & 6501  & 718.12 & 4171  & 209.1 \\ 
Pre-selection     & $E_T^{miss} > 30$ GeV             & 147.7 & 5617  & 505.25 & 526.3 & 181.6 \\
                  & $Z \rightarrow \tau\tau$ Rejection & 145.8 & 5215  & 485.12 & 164.2 & 150.4 \\
                  & Jet Veto                          & 61.80 & 14.84 & 238.35 & 31.91 & 76.12 \\
                  & b-veto                            & 61.56 &  6.85 & 237.87 & 30.76 & 76.12 \\ \hline
Signal region     & $\Delta\phi < 1.575$, $M_T < 600$ GeV & 50.6$\pm$2.5 & 2.3$\pm$1.6 & 85.4$\pm$2.7 & $<$ 1.7 & 38$\pm$38 \\ \hline
Control region    & $\Delta\phi > 1.575$, $M_T < 600$ GeV & 10.9$\pm$1.1 & 4.6$\pm$2.3 & 151.9$\pm$3.6 & 30.8$\pm$4.2 & 38$\pm$38 \\ \hline
\end{tabular}
\label{cut_flow_h0j}
\end{center}
\end{table*}

Events are required to pass at least one of the ATLAS single-lepton or double-lepton triggers~\cite{atlas}. 
The following notation is used to label different trigger items:
e (electron), EM (electromagnetic), MU or mu (muon), I or i (isolated).
As an example 2EM15I means two isolated electromagnetic objects with transverse energy $E_T > 15$ GeV. 
The Level-1 (L1) trigger menus used here are 2EM15I, EM25I, EM60I, MU20 and MU40. 
For the Level-2 (L2) trigger and Event Filter (EF), events are required to pass the e25i, 2e15i, e60 or mu20i triggers. 
For the $e\mu$ channel, the trigger efficiency for L1 is 99.0\%; for L2 it is 96.7\%, and for EF it is 95.2\%. 
The trigger efficiency for signal is quite high and the trigger efficiency does not distort the shapes
of the kinematic variables of interest in the signal in a significant way. Table~\ref{cut_flow_h0j}
does not include the trigger efficiency.

\begin{figure}[h]
\begin{center}
\includegraphics [width=0.45\textwidth] {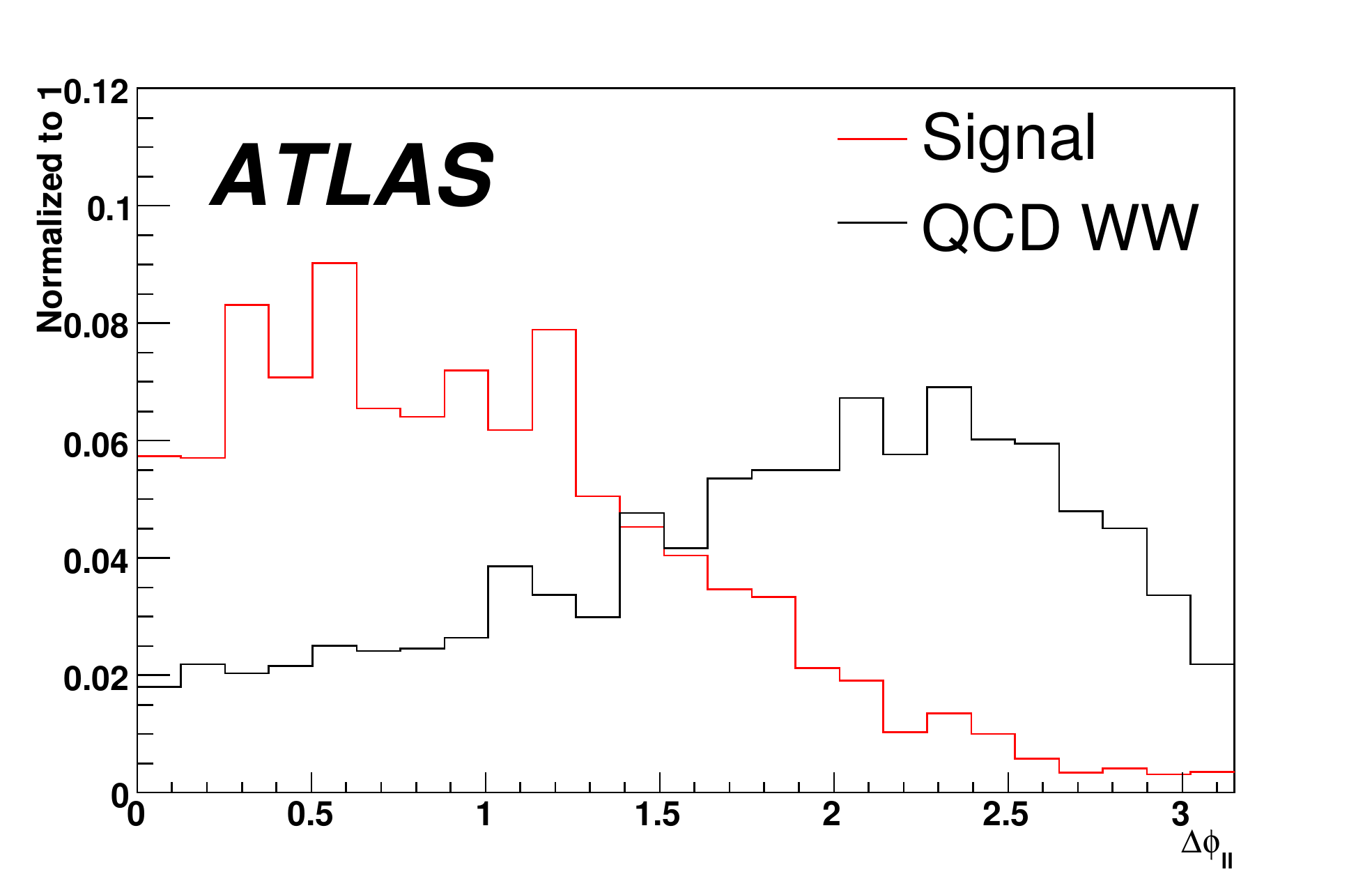}
\caption{Transverse opening angle $\Delta \phi_{\ell\ell}$ of two leptons after
preselection cuts.}
\label{fig:higgs_dphi}
\end{center}
\end{figure}

\begin{figure}[h]
\begin{center}
\includegraphics [width=0.45\textwidth] {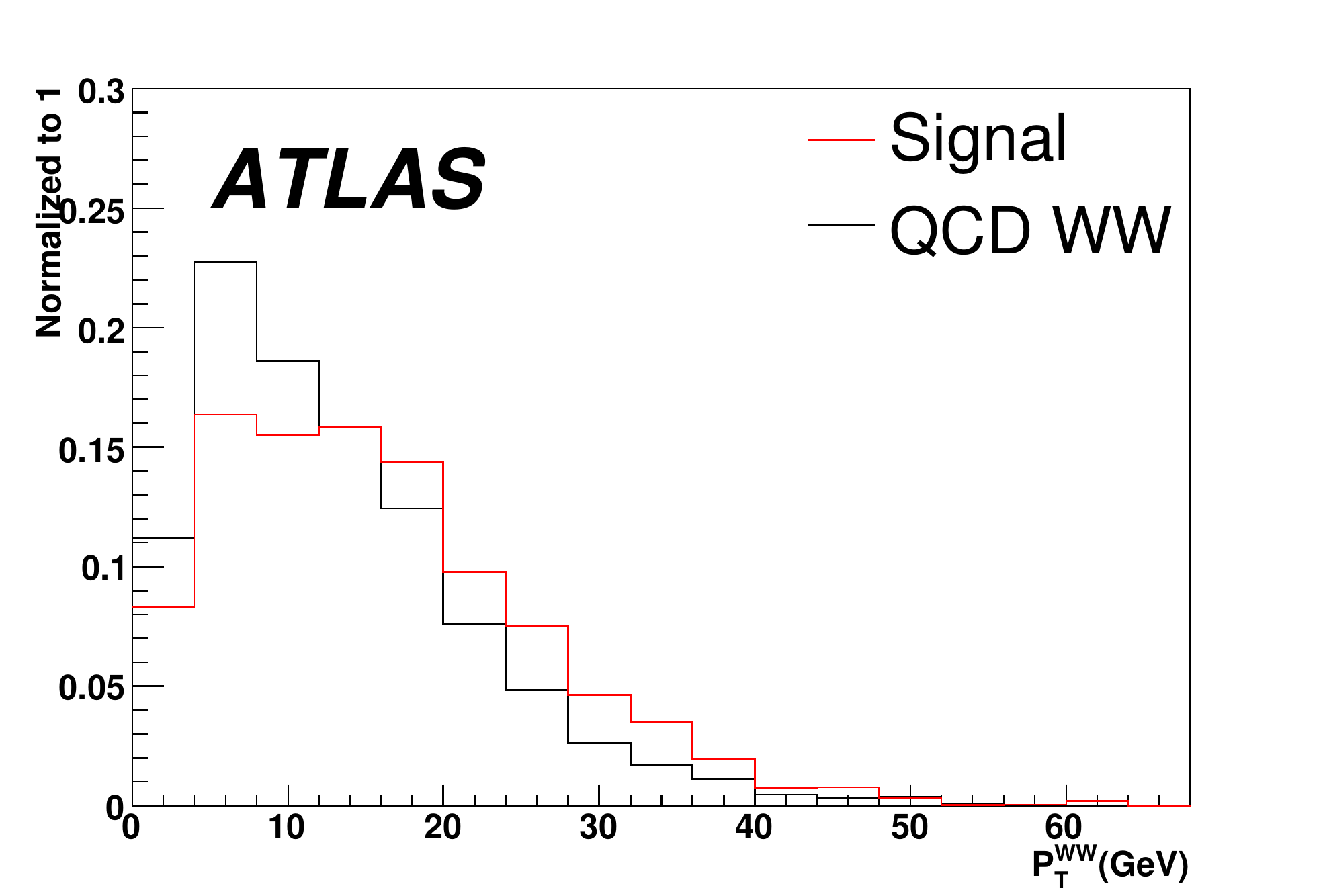}
\caption{Transverse momentum $p_T^{WW}$ of the WW system after preselection cuts.}
\label{fig:higgs_ptww}
\end{center}
\end{figure}

\begin{figure}[h]
\begin{center}
\includegraphics [width=0.45\textwidth] {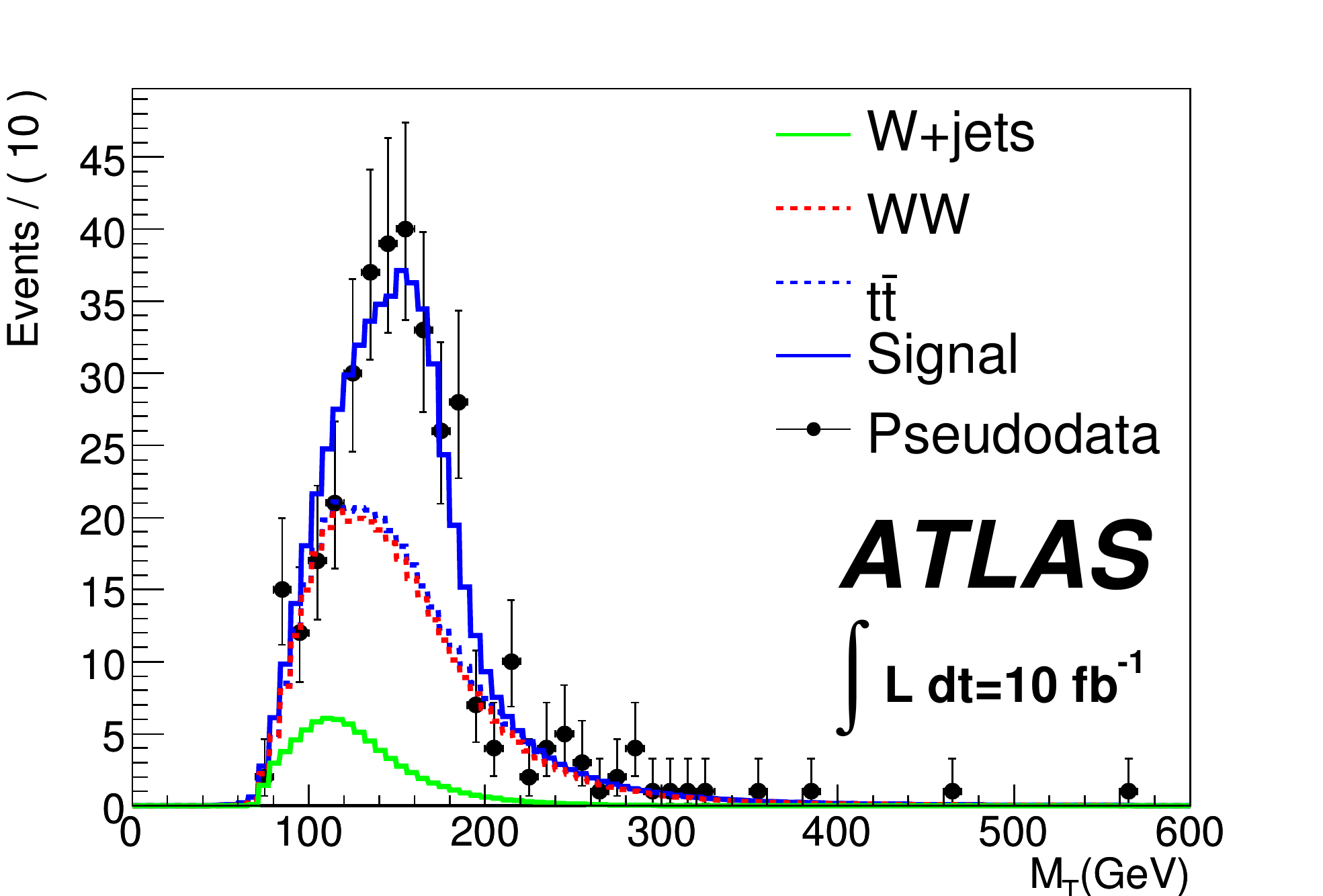}
\caption{Transverse mass $M_T$ for events with $\Delta \phi_{\ell\ell} < 1.575$ and $p_T^{WW} > 20$ GeV
assuming a SM Higgs boson with $M_H = 170$ GeV and 10 fb$^{-1}$ integrated luminosity.}
\label{fig:higgs_mt}
\end{center}
\end{figure}

After the basic event selection, the signal to background ratio is about 1:4. This analysis focuses
on three variables:
\begin{itemize}
\item the transverse opening angle $\Delta \phi_{\ell\ell}$; a cut on this variable exploits differences
in the spin correlations in the WW system in the Higgs boson signal and the WW background (Figure~\ref{fig:higgs_dphi}).
\item the transverse momentum of the WW system, $p_T^{WW}$, which tends to be slightly larger for
signal than for WW background because gluon-initiated processes tend to have more initial-state
radiation than quark-initiated processes (Figure~\ref{fig:higgs_ptww}).
\item the transverse mass which is defined as $M_T = \sqrt{(E_T^{\ell\ell} + E_T^{miss})^2 - (p_T^{\ell\ell} + p_T^{miss})^2}$ (Figure~\ref{fig:higgs_mt}).
\end{itemize}

\subsection{Fitting Algorithm}

A 2-dimensional fit of transverse mass ($M_T$) and $p_T^{WW}$ in two bins of the dilepton opening angle
$\Delta \phi_{\ell\ell}$ in the transverse plane has been implemented. After the preselection cuts and
the additional requirement that $M_T < 600$ GeV is applied, the remaining events are separated into
two subsamples, one with $\Delta \phi_{\ell\ell} < 1.575$ (signal region) and the other with 
$\Delta \phi_{\ell\ell} > 1.575$ (control region), as listed in Table~\ref{cut_flow_h0j}.

The top background is estimated with the help of b-tagged control samples with the same kinematic
cuts as the signal-enriched and background-enriched regions. The $t\bar{t}$ cross section is estimated
based on Monte Carlo samples that use MC@NLO to model the top background. Standalone fits are performed
on the b-tagged control samples before the fit to the b-vetoed regions begins, the top background in 
the b-vetoed regions is estimated by extrapolating both the shape and the normalization from the b-tagged
region to the b-vetoed region based on ratios obtained from MC@NLO.

The $Z \rightarrow \tau\tau$ background is normalized and its shape is determined by studying a sample of
$Z \rightarrow \mu\mu$ events taken from real data, where the reconstructed muons are replaced by simulated taus.
Two muons events with a dimuon invariant mass between 82 and 98 GeV are selected, and the same jet veto is applied 
to selected events. After event selection cuts are applied, the effective cross section of $Z \rightarrow \tau\tau$ 
background is about 250 fb. A standalone fit to these ``data-Monte-Carlo'' events is performed to determine
the shape and normalization of the $Z \rightarrow \tau\tau$ background; in the final fit, the shape 
parameters obtained from the fit to control sample are fixed and the normalization is rescaled by a factor 
that is assumed in this study to be well-predicted.

W+jets is one of the main sources of fake backgrounds for the dilepton channels, it is crucial to 
achieve a good rejection against this background. The average jet fake as electron rate is 
$(1.7 \pm 0.2) \times 10^{-4}$ before isolation and $(6.7 \pm 1.5) \times 10^{-5}$ after electron isolation.
The corresponding jet fake as muon rate is $(1.7 \pm 0.6) \times 10^{-5}$ after isolation cuts are applied.
For W+jets background estimation, due the limited size of the available W+jets Monte Carlo sample,
the shape of the transverse mass and $p_T^{WW}$ distributions are taken from a set of events with 
loosen isolation and shower shape cuts. Further detailed studies are needed to estimate W+jets background.

Once the fits to the control samples are completed, a simultaneous fit to the two $\Delta \phi_{\ell\ell}$ bins in 
the b-vetoed region is performed. A few of the parameters that described the shape of the transverse mass and
$p_T^{WW}$ distributions of the WW background are allowed to float in the fit. The normalizations of the WW 
background are free to float independently. However, we add a penalty term of the form  
$(R_{fit} - R_{true}^2) / \sigma_R^2$, where $R_{fit}$ is the ratio of the best-fit number of WW background
events in the small-$\Delta \phi_{\ell\ell}$ region over the number in the large large-$\Delta \phi_{\ell\ell}$
region, $R_{true}$ is the Monte Carlo prediction of the ratio taken from the central-value calculation, and 
$\sigma_R$ is the uncertainty in the prediction of $R_{true}$, taken to be 10\%.
\begin{figure}[h]
\begin{center}
\includegraphics [width=0.45\textwidth] {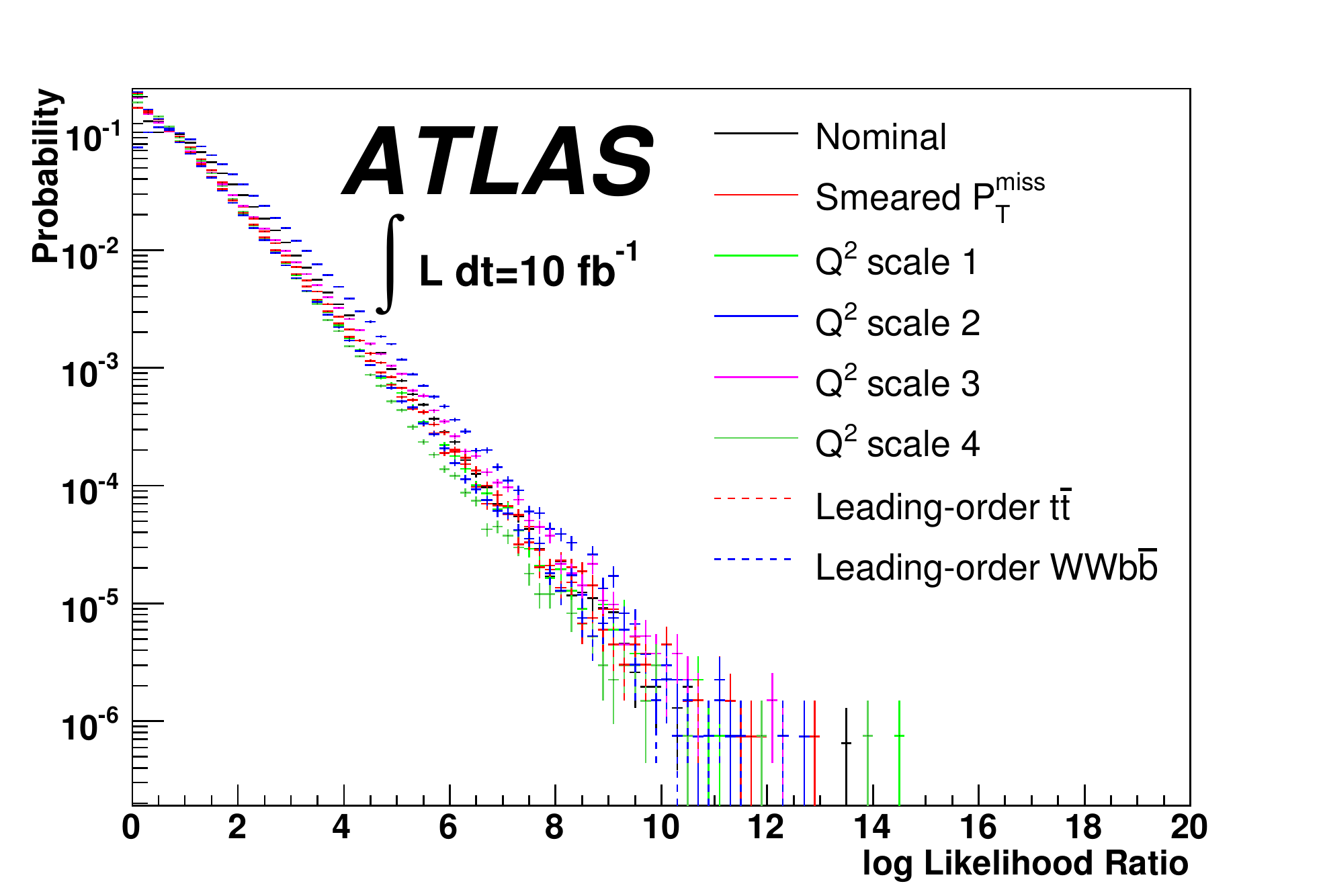}
\caption{The log Likelihood Ratio distributions for background-only Monte Carlo 
in H + 0j, $H \rightarrow WW \rightarrow e\nu\mu\nu$ assuming 10 fb$^{-1}$.}
\label{fig:higgs_syst_bkgd}
\end{center}
\end{figure}
\begin{figure}[h]
\begin{center}
\includegraphics [width=0.45\textwidth] {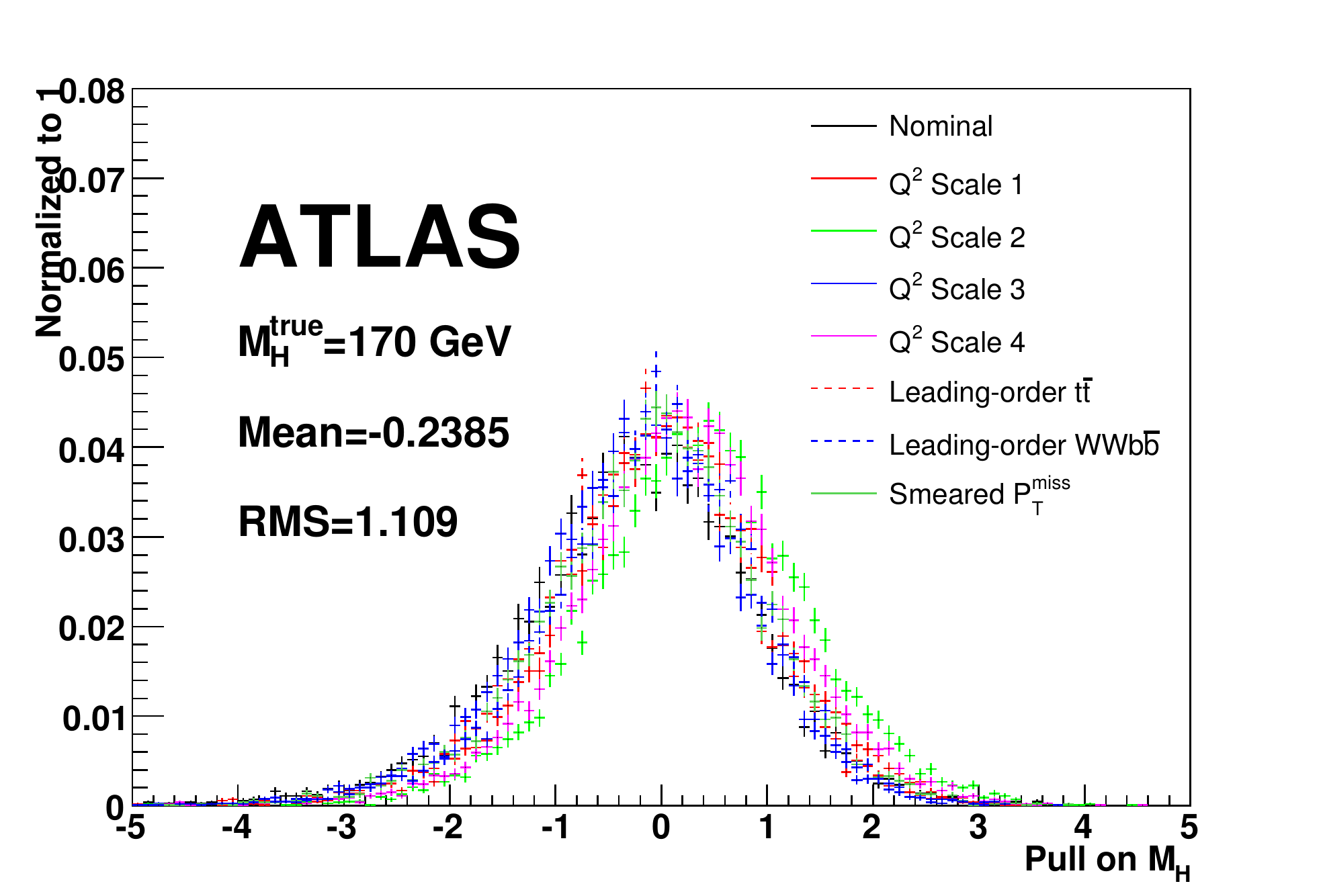}
\caption{The pull distributions for $M_H = 170$ GeV for Higgs boson signal and background Monte Carlo 
in H + 0j, $H \rightarrow WW \rightarrow e\nu\mu\nu$ assuming 10 fb$^{-1}$.}
\label{fig:higgs_syst_signal}
\end{center}
\end{figure}

In order to demonstrate the robustness of the fit against systematic uncertainties, top Monte Carlo has been
used to compute the sampling distributions for the Likelihood Ratio in several scenarios where the ``true''
probability distribution has been distorted to model various sources of systematic error. Seven distorted scenarios
are considered: four altered $Q^2$ scale choices (factorization and renormalization scales raised and 
lowered by factors of 8), two alternative top background models (based on leading-order $pp \rightarrow WWbb$
and leading-order $pp \rightarrow tt \rightarrow WWbb$), and one alternative model of all irreducible 
backgrounds where the $x$ and $y$ components of $E_T^{miss}$ have been independently smeared by 5 GeV each.
Figure~\ref{fig:higgs_syst_bkgd} shows the Likelihood Ratio distributions for background-only outcomes and 
Figure~\ref{fig:higgs_syst_signal} represents the distributions of pulls of the fitted Higgs boson mass for
signal plus background with a true Higgs boson mass of 170 GeV.

\begin{figure}[h]
\begin{center}
\includegraphics [width=0.45\textwidth] {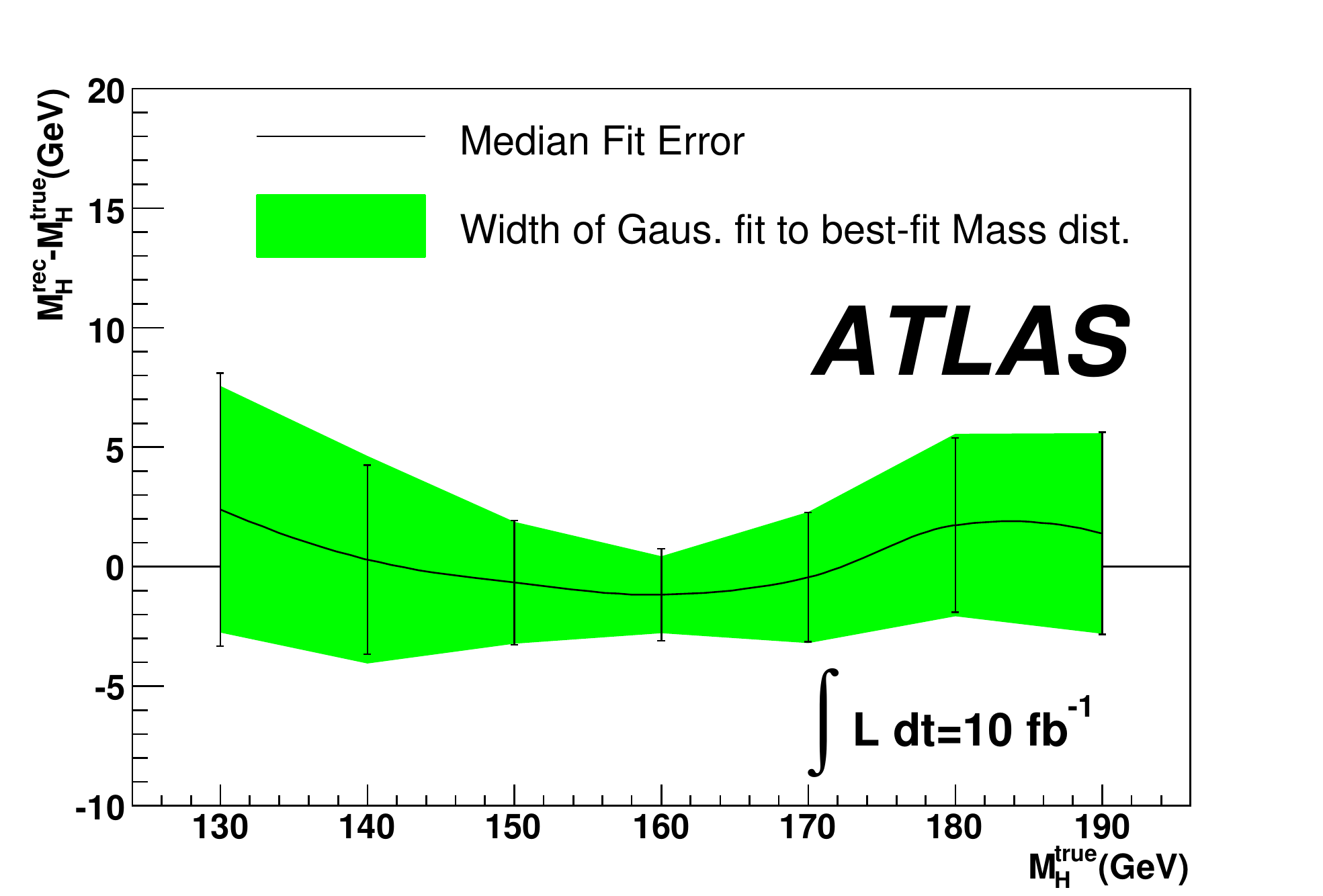}
\caption{The linearity of the Higgs boson mass determination for H + 0j, $H \rightarrow WW \rightarrow e\nu\mu\nu$ 
at 10 fb$^{-1}$ integrated luminosity.}
\label{fig:higgs_mass}
\end{center}
\end{figure}

The linearity of the mass determination as a function of the true Higgs boson mass is shown in Figure~\ref{fig:higgs_mass}.
The line shows the mean of a Gaussian fit to the region around the peak of the distribution of best-fit Higgs boson masses
in the Monte Carlo sample for the case of nominal detector performance. The green band shows the width of the Gaussian
fit and is a direct measure of the variability of the mass estimate on repetition of the experiment, the error bars
show the median fit error. The typical variability of the mass determinations with 10 fb$^{-1}$ integrated
luminosity are 5.2 GeV, 1.6 GeV and 4.2 GeV at $M_H = 130, ~160$ and $190$ GeV, respectively. 

\begin{figure}[h]
\begin{center}
\includegraphics [width=0.45\textwidth] {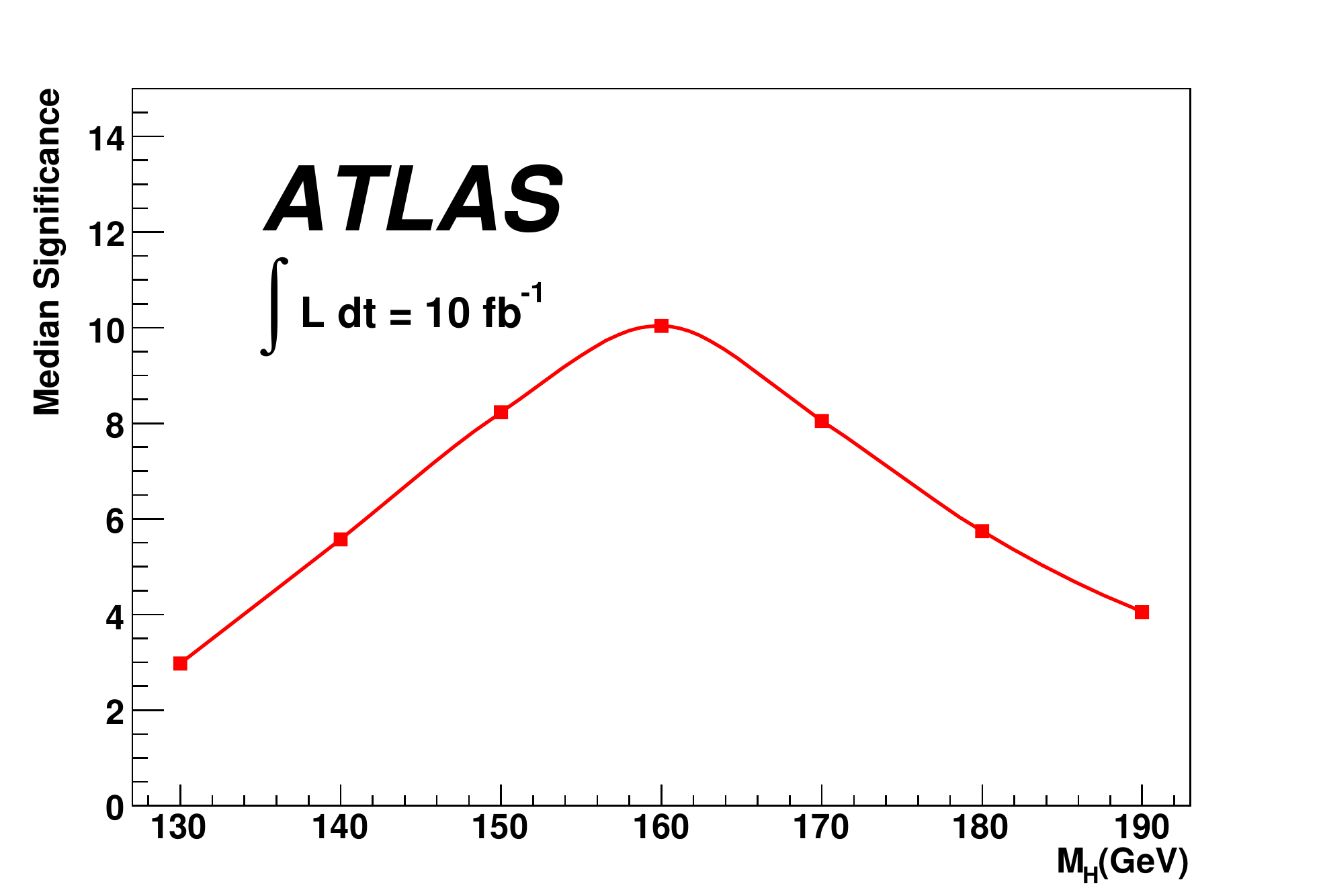}
\caption{The expected Higgs boson detection significance for the H + 0j, $H \rightarrow WW \rightarrow e\nu\mu\nu$
at 10 fb$^{-1}$ integrated luminosity.}
\label{fig:higgs_significance}
\end{center}
\end{figure}

The $H \rightarrow WW \rightarrow e\nu\mu\nu$ is the
most promising channel for Higgs boson masses near the WW threshold.
The expected SM Higgs boson detection significance assuming 10 fb$^{-1}$ integrated luminosity is shown 
in Figure~\ref{fig:higgs_significance}. 



\section{Leptonic W Pair Production with Two Hard Jets}

For SM Higgs boson produced through VBF production mechanism,
the final state has two energetic leptons from both W bosons leptonic decays
associated with two hard jets. The distinctive characteristics of VBF Higgs boson signal
include:
\begin{itemize}
\item The two jets arising from struck quarks tend to be the highest-$p_T$ jets in the events, 
and they tend to be well-separated in pseudo-rapidity;
\item they tend to have a large invariant mass;
\item there is very little jet activity in the region between the two tagging jets.
\end{itemize}
\begin{figure*}[htbp]
\begin{center}
\includegraphics [width=0.8\textwidth] {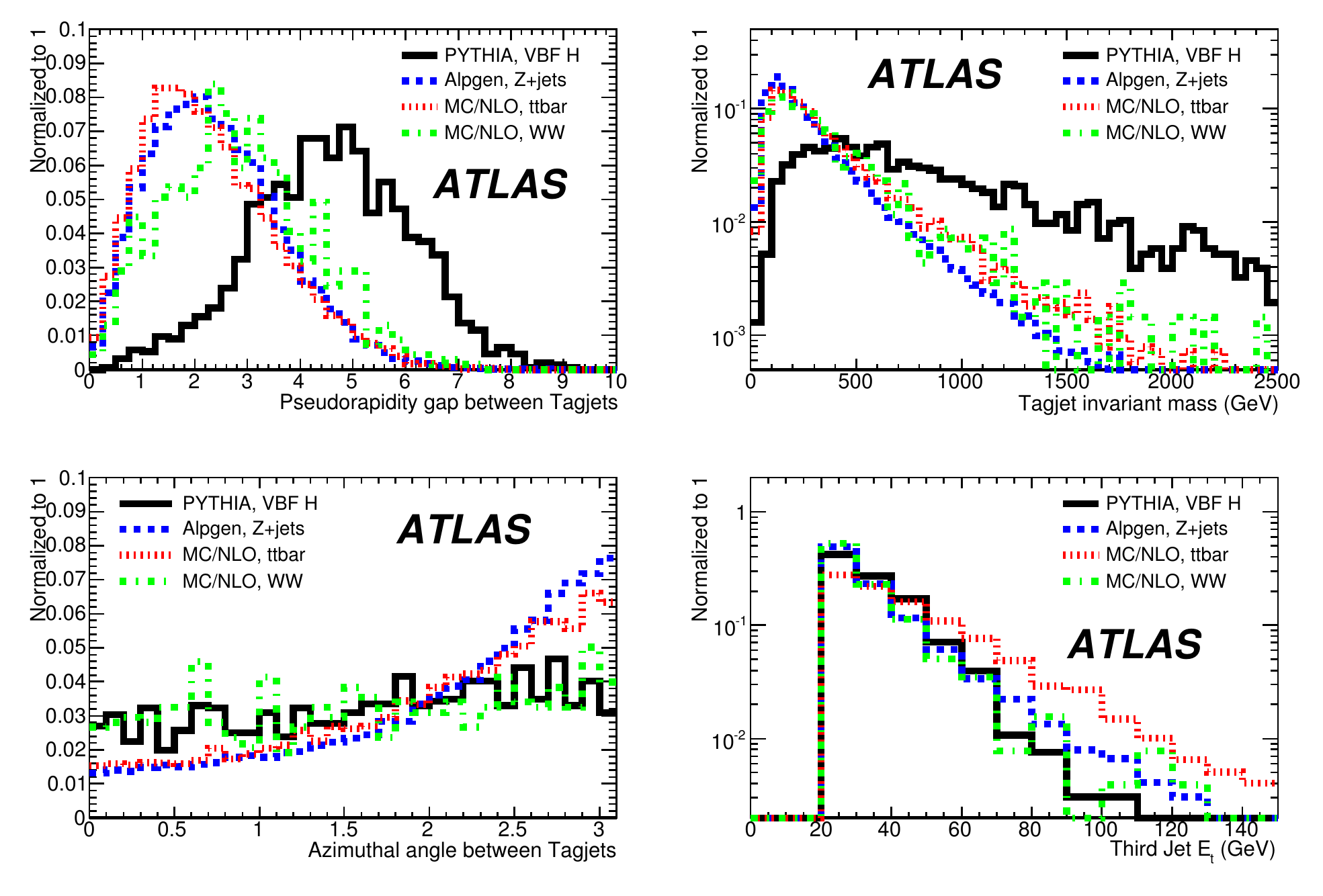}
\caption{Pseudo-rapidity gap between two tagging jets(top left),
invariant mass distribution of two tagging jets(top right).
azimuthal angle gap between two tagging jets(bottom left) and
$E_t$ of the third jet(bottom right).}
\label{fig:higgs_h2j}
\end{center}
\end{figure*}
\begin{table*}[t]
\begin{center}
\caption{Cut flows(in fb) for $M_H=170$ GeV in the H + 2j, $H \rightarrow WW \rightarrow e\nu\mu\nu$ channel.}
\begin{tabular}{|l||c|c|c|c|c|}
\hline Selection Cuts             &~~VBF H (170 GeV)~~ & ~~$t\bar{t}$~~  & ~~ WW+jets ~~ & ~~ $Z\rightarrow \tau\tau$ ~~ & ~~W+jets ~~\\ \hline
Lepton Selection                  & 30.20 &  8317   & 838.96 & 2096   & 1323  \\ 
Forward Jet Tagging               & 17.27 &   946.6 &  32.77 & 79.3   & 31.83 \\
Leptons Between Jets              & 16.47 &   617.8 &  22.92 & 55.13  & 27.91 \\
$Z \rightarrow \tau\tau$ Rejection& 15.68 &   561.8 &  21.20 & 39.03  & 27.91 \\
$E_T^{miss}$,$M_T$,$m_T^{\ell\ell\nu}$& 12.78&425.9 &  15.28 & 0      & 13.96 \\
b-veto                            & 12.67 &  206.72 &  -     & -      & - \\ \hline

signal box, b-jet veto            & 9.28$\pm$0.27 & 28.5$\pm$5.7 & 4.75$\pm$0.30 & - & 4.3$\pm$4.3 \\  
signal box, no b-jet veto         & 9.65          & 114.2        & 4.99          & - & 6.07 \\ \hline

Control box, b-jet veto           & 3.02$\pm$0.15 & 89$\pm$10    & 9.78$\pm$0.43 & - & 7.9$\pm$5.0 \\
Control box, no b-jet veto        & 3.13          & 311.7        & 10.28         & - & 7.89 \\ \hline

\end{tabular}
\label{cut_flow_h2j}
\end{center}
\end{table*}

Figure~\ref{fig:higgs_h2j} shows the distribution of the pseudo-rapidity gap between two tagging jets (top left),
invariant mass of two tagging jets (top right), azimuthal angle between two tagging jets (bottom left) and
transverse energy of the third jet (bottom right). Where black solid histograms represent VBF Higgs boson signal with
$M_H = 170$ GeV, green dash-dotted histograms mean WW background, red dotted histograms are $t\bar{t}$ and 
blue dashed histograms are Z+jets background.

The following cuts have been applied to selected H + 2j, $H \rightarrow WW \rightarrow \ell\nu\ell\nu$ events.
\begin{itemize}
\item Two isolated leptons with $p_T > 15$ GeV
\item Missing transverse energy $E_T^{miss} > 30$ GeV
\item At least two jets with $p_T > 20$ GeV and $|\eta| < 4.8$
\item The two jets with highest transverse momentum are required to be in opposite hemispheres with
$\Delta \eta(jet1,jet2) > 3$
\item Require that both leptons are between the two leading jets in pseudo-rapidity
\item Reject event which has $|M_{\tau\tau} - M_Z| < 25$ GeV
\item Require transverse mass $50 < M_T < 600$ GeV
\item Require transverse mass $m_T^{\ell\ell\nu} > 30$ GeV, 
$m_T^{\ell\ell\nu} = \sqrt{2 P_T(\ell\ell) P_T^{miss} \cdot (1-cos\Delta\phi)}$
where $\Delta\phi$ is the angle between the di-lepton vector and the $P_T^{miss}$ vector in the transverse plane.
\item To suppress b-related background, b-veto cuts are applied on both tagging jets
\end{itemize}
\begin{figure}[htb]
\begin{center}
\includegraphics [width=0.45\textwidth] {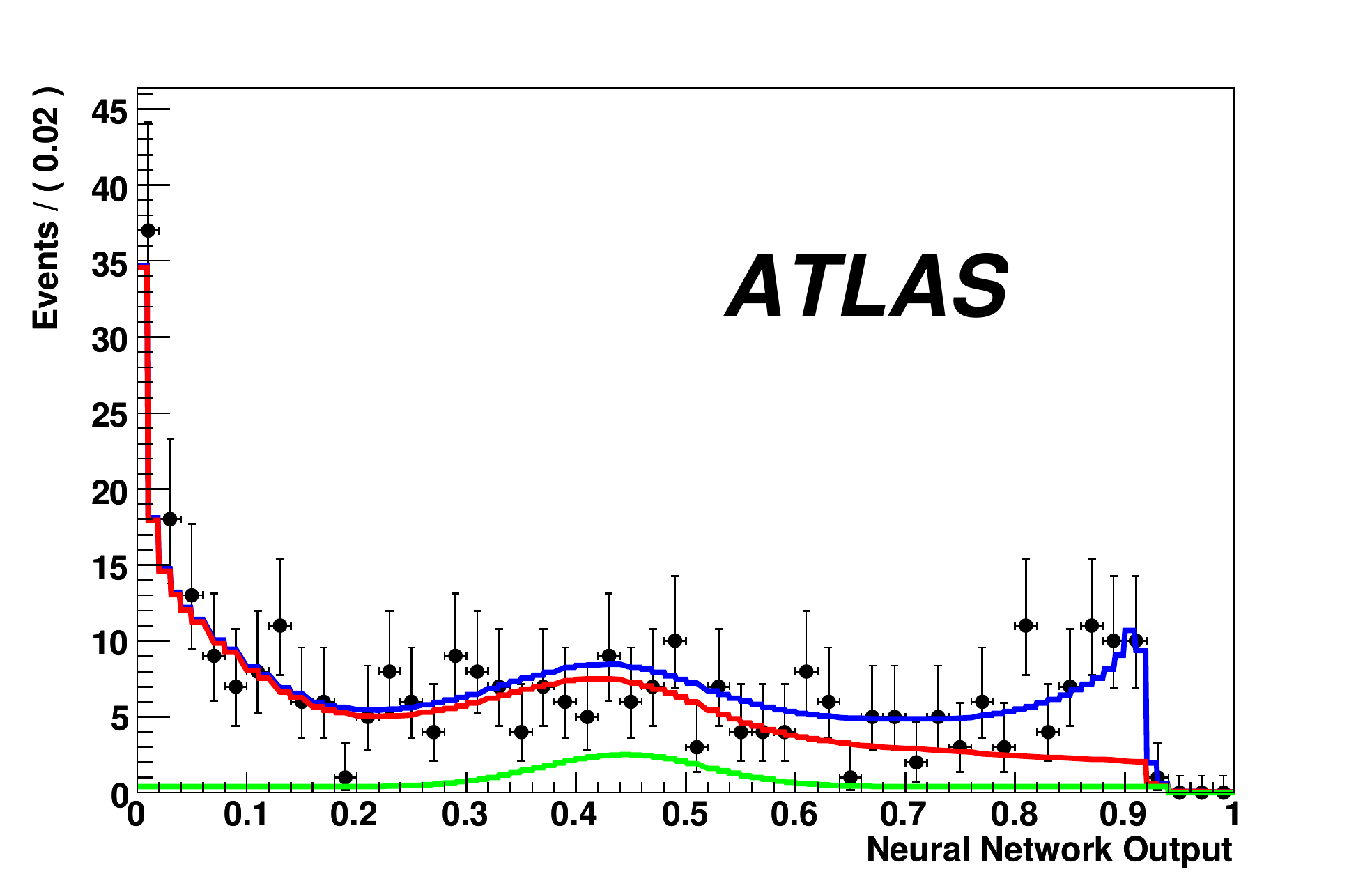}
\caption{The Neural Network output distribution in the signal box,
for the 170 GeV H + 2j, $H \rightarrow WW \rightarrow e\nu\mu\nu$
at 10 fb$^{-1}$ integrated luminosity.}
\label{fig:higgs_nnout_h2j}
\end{center}
\end{figure}

\begin{figure}[htb]
\begin{center}
\includegraphics [width=0.45\textwidth] {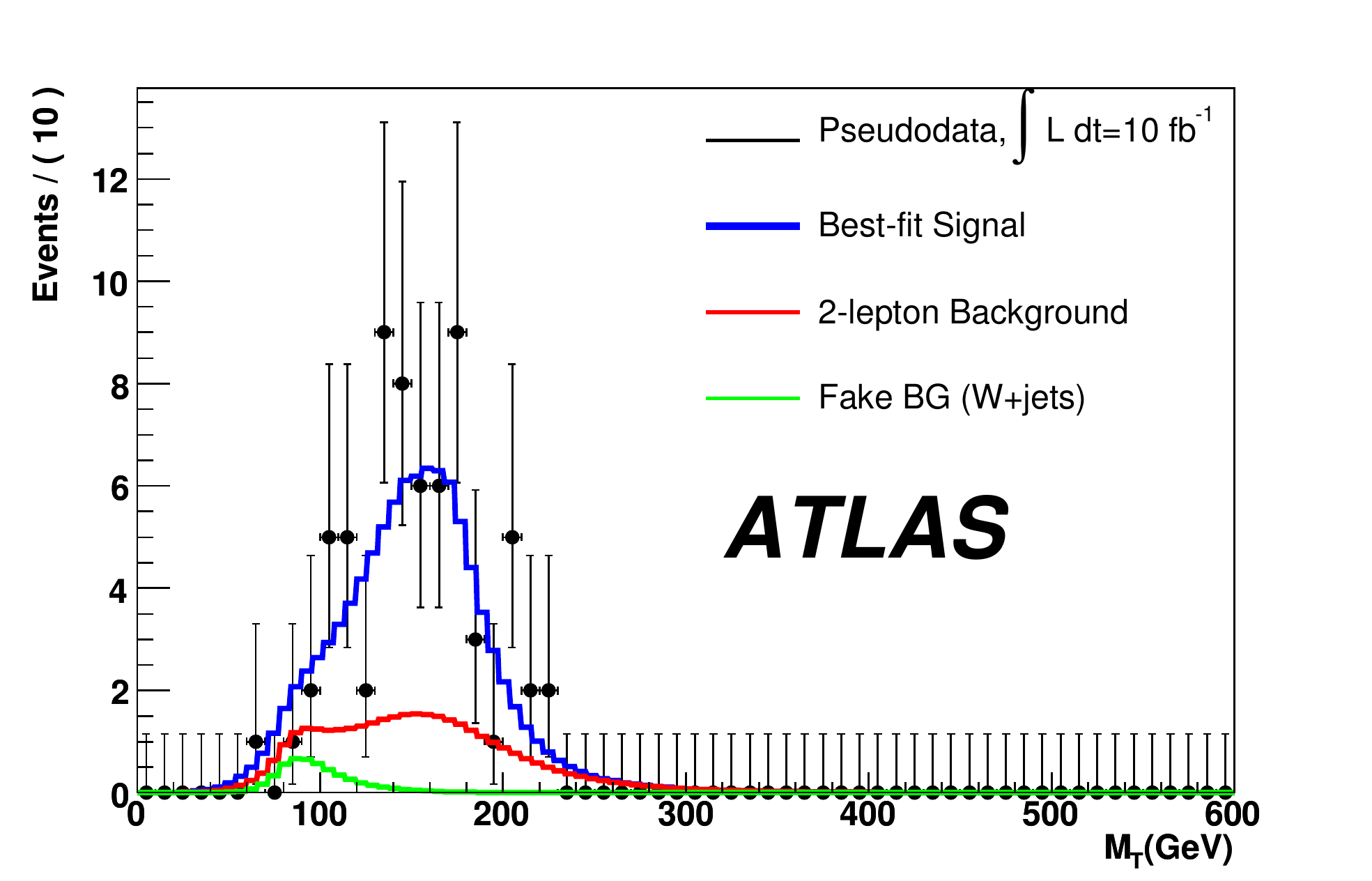}
\caption{The transverse mass distribution for events in the signal box
with Neural Network output larger than 0.8 for the 170 GeV 
H + 2j, $H \rightarrow WW \rightarrow e\nu\mu\nu$ at 10 fb$^{-1}$ integrated luminosity.}
\label{fig:higgs_mt_h2j}
\end{center}
\end{figure}

\begin{figure}[htb]
\begin{center}
\includegraphics [width=0.45\textwidth] {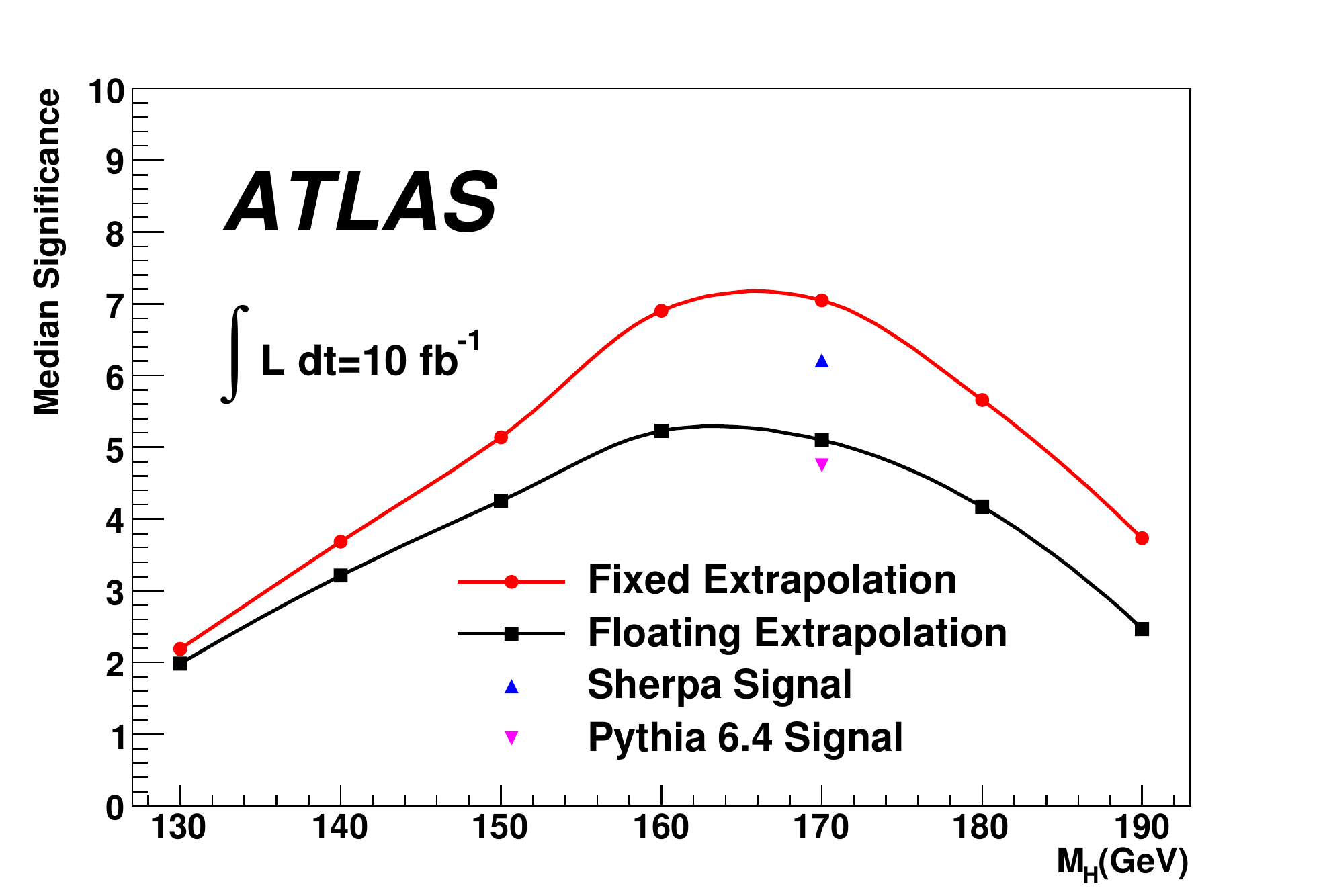}
\caption{The expected Higgs boson detection significance for the H + 2j, $H \rightarrow WW \rightarrow e\nu\mu\nu$
at 10 fb$^{-1}$ integrated luminosity.}
\label{fig:higgs_significance_h2j}
\end{center}
\end{figure}

Table~\ref{cut_flow_h2j} shows the cut flows (in fb) for $M_H = 170$ GeV in the H + 2j, $H \rightarrow WW \rightarrow e\nu\mu\nu$ channel.
If the event has $\Delta \phi_{\ell\ell} < 1.5$ and $\Delta \eta_{\ell\ell} < 1.4$, it lies in the signal box;
otherwise, it lies in the control region. The trigger efficiency of 99.0\% after Level 1, 96.8\% after Level 2,
and 94.5\% afte the Event Filter. 

After the preselection, a four-variable Neural Network is used to further enhance the separation between 
the signal and the background. The inputs to the Neural Network are:
\begin{itemize}
\item $\Delta \eta(jet1,jet2)$, the pseudo-rapidity gap between two tagging jets
\item $M_{jj}$, the invariant mass of two tagging jets
\item the transverse momentum of the leading non-tagging jet in the region $|\eta|<3.2$
\item $\eta^* = \eta_3 - (\eta_1 + \eta_2)/2$, the pseudo-rapidity gap between two tagging jets and the third
non-tagging jet.
\end{itemize}

The Neural Network output distribution in the signal box is shown in Figure~\ref{fig:higgs_nnout_h2j}.
The transverse mass distribution for events in the signal box with Neural Network output larger than 0.8
is shown in Figure~\ref{fig:higgs_mt_h2j}, where black dots with error bar show Pseudo-data, 
blue curve shows best-fit Higgs boson signal, red curve represents
2-lepton background and green curve is faked background from W+jets.

The fit is a two-dimensional fit to the Neural Network output and transverse mass distributions.
Both the signal model and the background model are uncorrelated product probability density functions (PDFs).
The Neural Network output distribution for the background in the signal box is taken to be
the same as the distribution in the control region, but it is multiplied by a linear extrapolation
factor. Apart from the slope of this extrapolation factor, all parameters governing the shape
of the Neural Network output distribution in the two regions are required to be the same.
The expected significance is shown as a function of the true Higgs boson mass in Figure~\ref{fig:higgs_significance_h2j}.

\section{Further Improvements}

We expect further improvements for Higgs boson detection sensitivity. The analysis presented above
only includes the $e\nu\mu\nu$ channel with jet-veto or two forward jet tagging in events. 
Including the $e\nu e\nu$ and $\mu\nu \mu\nu$ channels and adding dilepton events with 1 jet
in final states will certainly increase the detection sensitivity in ATLAS. 
Based on LEP and Tevatron Higgs boson search experiences, the analysis can be carried out by 
using mativariate techniques, such as  Artificial Neural Networks (ANN) and 
Boosted Decision Trees (BDT)~\cite{bdt}, which would further improve the detection
sensitivities. ATLAS physics group has extensively explored these advanced analysis
methods~\cite{atlas} and fully developed the tools that will be used in Higgs boson search program with
LHC collision data.

\begin{figure}[htb]
\begin{center}
\includegraphics [width=0.45\textwidth] {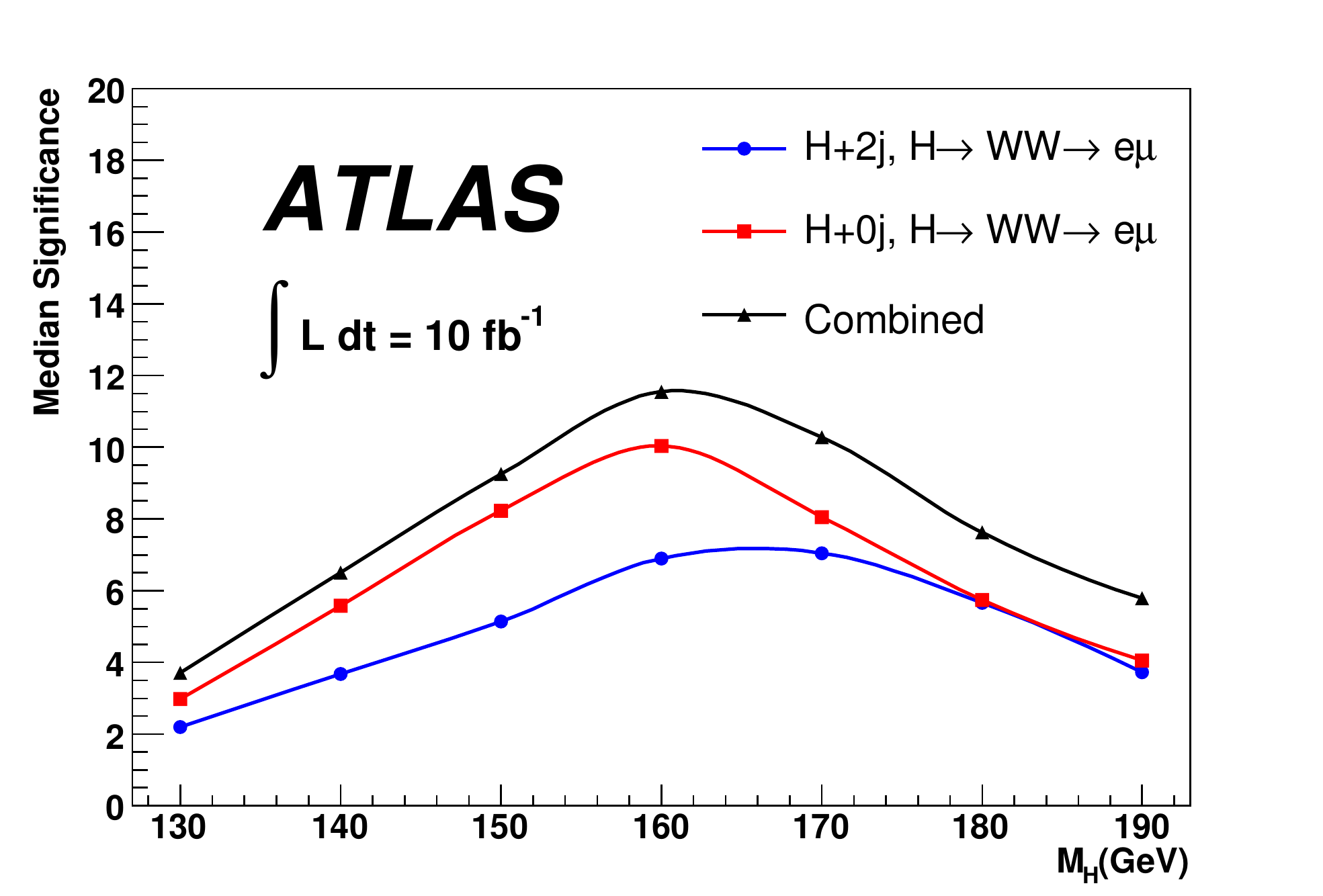}
\caption{The expected Higgs boson detection significance for the H + 0j and H + 2j, 
$H \rightarrow WW \rightarrow e\nu\mu\nu$ at 10 fb$^{-1}$ integrated luminosity.}
\label{fig:higgs_significance_combined}
\end{center}
\end{figure}
\begin{figure}[htb]
\begin{center}
\includegraphics [width=0.45\textwidth] {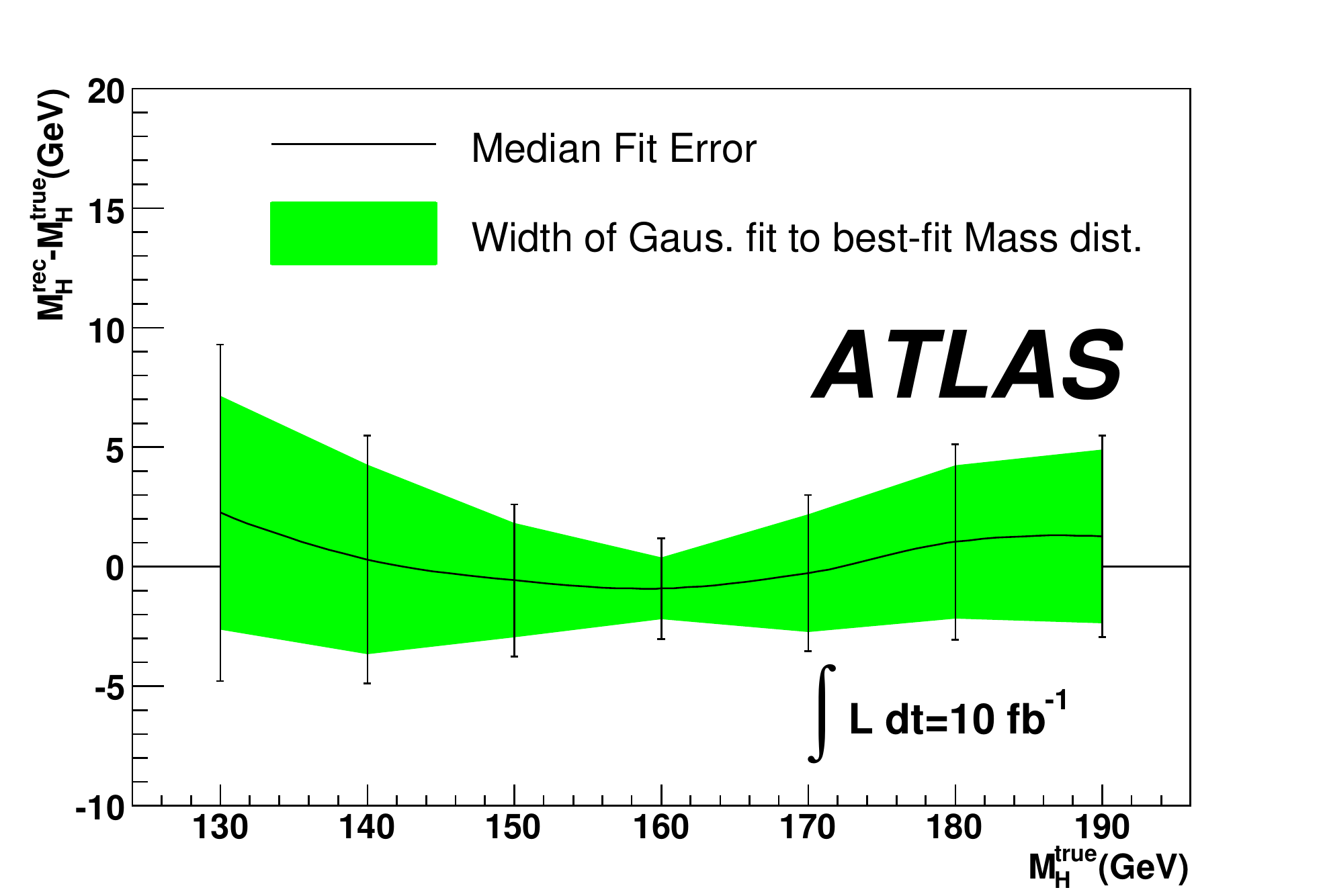}
\caption{The linearity of the mass determination for the combined fit of H + 0/2j, 
$H \rightarrow WW \rightarrow e\nu\mu\nu$ at 10 fb$^{-1}$ integrated luminosity.}
\label{fig:higgs_mass_combined}
\end{center}
\end{figure}
\section{Conclusions}

The prospects for SM Higgs boson searches in the WW decay mode have been studied using
a realistic model of the ATLAS detector. The H + 0j, $H \rightarrow WW \rightarrow e\nu\mu\nu$
channel is very promising for Higgs boson masses in the region around the WW threshold. 
With 10 fb$^{-1}$ of integrated luminosity, one would expect to be able to achieve a $5\sigma$
discovery with the $H \rightarrow WW \rightarrow e\nu\mu\nu$ channel alone if there a
SM Higgs boson with a mass between $\sim$ 140 GeV and $\sim$ 185 GeV. A measurement of the mass
of the Higgs boson at 10 fb$^{-1}$ integrated luminosity would have a precision of less than 2 GeV
for Higgs mass boson of 160 GeV, or a precision of less than 4 GeV for Higgs boson with mass of 140 GeV.
The H + 2j, $H \rightarrow WW \rightarrow e\nu\mu\nu$ channel has a smaller event rate than the H + 0j
channel. With 10 fb$^{-1}$ of integrated luminosity, we expect to reach $5\sigma$ discovery if
SM Higgs boson mass between 150 GeV and 180 GeV. 
The combined mass determination for the combined fit of H + 0/2j is shown in 
Figure~\ref{fig:higgs_mass_combined}.
The corresponding combined significance as a function of Higgs boson mass 
is shown in Figure~\ref{fig:higgs_significance_combined}.

\begin{acknowledgments}
The author would like to express gratitude to the ATLAS Collaboration for excellent
work on the Monte Carlo simulation and the software package for physics analysis.
The ATLAS Higgs working group deserves a special thank for producing the results
presented in this paper.
The author is supported by the Department of Energy (DE-FG02-95ER40899) of the United States.
\end{acknowledgments}

\bigskip 

\end{document}